\documentclass[11pt,preprint]{aastex}
\shorttitle{C$^{14}$N/C$^{15}$N IN DIFFUSE CLOUDS}
\shortauthors{RITCHEY ET AL.}

\begin{document}
\title{The C$^{14}$N/C$^{15}$N Ratio in Diffuse Molecular Clouds\footnote{Based on observations made with the Very Large Telescope of the European Southern Observatory, Paranal, Chile, under programs 065.I-0526, 071.C-0367, 071.C-0513, 076.C-0431, and 092.C-0019.}}
\author{A.~M.~Ritchey\altaffilmark{1}, S.~R.~Federman\altaffilmark{2}, and D.~L.~Lambert\altaffilmark{3}}
\altaffiltext{1}{Department of Astronomy, University of Washington, Seattle, WA 98195, USA; aritchey@astro.washington.edu}
\altaffiltext{2}{Department of Physics and Astronomy, University of Toledo, Toledo, OH 43606, USA; steven.federman@utoledo.edu}
\altaffiltext{3}{W.~J.~McDonald Observatory and Department of Astronomy, University of Texas at Austin, Austin, TX 78712, USA; dll@astro.as.utexas.edu}

\begin{abstract}
We report the first detection of C$^{15}$N in diffuse molecular gas from a detailed examination of CN absorption lines in archival VLT/UVES spectra of stars probing local diffuse clouds. Absorption from the C$^{15}$N isotopologue is confidently detected (at $\gtrsim4\sigma$) in three out of the four directions studied and appears as a very weak feature between the main $^{12}$CN and $^{13}$CN absorption components. Column densities for each CN isotopologue are determined through profile fitting, after accounting for weak additional line-of-sight components of $^{12}$CN, which are seen in the absorption profiles of CH and CH$^+$ as well. The weighted mean value of C$^{14}$N/C$^{15}$N for the three sight lines with detections of C$^{15}$N is $274\pm18$. Since the diffuse molecular clouds toward our target stars have relatively high gas kinetic temperatures and relatively low visual extinctions, their C$^{14}$N/C$^{15}$N ratios should not be affected by chemical fractionation. The mean C$^{14}$N/C$^{15}$N ratio that we obtain should therefore be representative of the ambient $^{14}$N/$^{15}$N ratio in the local interstellar medium. Indeed, our mean value agrees well with that derived from millimeter-wave observations of CN, HCN, and HNC in local molecular clouds.
\end{abstract}


\section{INTRODUCTION}
Among the important (yet ill-understood) diagnostics of stellar nucleosynthesis and the subsequent chemical enrichment of the interstellar medium (ISM) is the $^{14}$N/$^{15}$N isotopic ratio. The production of $^{14}$N is thought to involve a combination of secondary sources, such as the cold CNO cycle in main sequence stars, and primary ones, such as Hot Bottom Burning in asymptotic giant branch (AGB) stars, while $^{15}$N is thought to be synthesized almost exclusively as a secondary nuclide, e.g., through the hot CNO cycle associated with nova outbursts. Various sources of $^{14}$N and $^{15}$N have been incorporated into models of Galactic chemical evolution (GCE; e.g., Tosi 1982; Romano \& Matteucci 2003; Kobayashi et al.~2011), which commonly predict that the $^{14}$N/$^{15}$N ratio should increase with Galactocentric radius. Millimeter-wave emission studies of nitrogen-bearing molecules and their isotopologues in warm dense clouds of the Galactic disk have indeed revealed such a trend (e.g., Dahmen et al.~1995; Wilson 1999; Adande \& Ziurys 2012). Adande \& Ziurys (2012), who examined millimeter observations of CN and HNC in 11 molecular clouds, and reevaluated previous observations of HCN, inferred a nitrogen isotopic ratio for the solar neighborhood of $290\pm40$ from the gradient seen in their data. 

Other determinations of the $^{14}$N/$^{15}$N ratio in the local ISM sometimes give quite different results, however, suggesting that the ratios may be affected by chemical fractionation. Hily-Blant et al.~(2013a, 2013b) found significant differences in the nitrogen isotopic ratios derived from observations of CN and HCN isotopologues in dark molecular cloud cores. An overall ratio of $500\pm75$ was obtained for CN, while HCN exhibited ratios between 140 and 360. The authors interpreted these results as evidence of fractionation and presented model calculations that they propose can explain the observations. Daniel et al.~(2013) studied a suite of nitrogen-bearing molecules and their isotopologues in the dark cloud Barnard 1, finding no evidence for nitrogen fractionation, which they suggest may be linked to the relatively high gas kinetic temperature ($T_{\mathrm{kin}}\approx17$~K) in the innermost part of the B1b cloud. Their molecular abundances were found to be consistent with a $^{14}$N/$^{15}$N ratio of $\sim$300 for the parental molecular cloud, in agreement with the Adande \& Ziurys (2012) result obtained for warmer sources in the local ISM. For comparison, Lucas \& Liszt (1998) obtained a nitrogen isotopic ratio of $237^{+27}_{-21}$ for diffuse molecular gas seen in absorption from HCN toward the extragalactic continuum source B0415$+$379.

In this Letter, we present a new approach to estimating the nitrogen isotopic ratio in the local ISM; we report the first detection of C$^{15}$N in diffuse molecular gas from an analysis of CN absorption lines in the optical spectra of background O and B-type stars. Specifically, we examine high quality optical spectra of HD~73882, HD~154368, HD~169454, and HD~210121, which have distances in the range $d$~$\approx$~0.2--1.1~kpc and color excesses of $E(B-V)$~$\approx$~0.3--1.1 mag. The diffuse molecular clouds probed by these sight lines have high enough kinetic temperatures ($T_{\mathrm{kin}}\gtrsim20$~K; e.g., Sonnentrucker et al.~2007) that chemical fractionation is not expected to be important. Furthermore, the amount of visual extinction through the clouds ($A_V\lesssim4$~mag; Valencic et al.~2004; Rachford et al.~2009) is small enough that selective photodissociation of N$_2$ should not play a significant role (Heays et al.~2014). These measurements have the potential therefore to yield the most accurate $^{14}$N/$^{15}$N ratio to date for material within the solar neighborhood ($d\lesssim1$~kpc).

\section{OBSERVATIONS AND DATA ANALYSIS}
The observations examined here were previously discussed by Ritchey et al.~(2011), who studied $^{12}$CN/$^{13}$CN and $^{12}$CH$^+$/$^{13}$CH$^+$ ratios along 13 lines of sight through diffuse molecular clouds. Upon further examination of those data, it was realized that for four sight lines with very strong CN absorption lines, a weak feature between the main $^{12}$CN and $^{13}$CN absorption components might be due to the C$^{15}$N isotopologue. The relevant data, acquired using the Ultraviolet and Visual Echelle Spectrograph (UVES) of the Very Large Telescope (VLT), were obtained from the European Southern Observatory (ESO) Science Archive Facility and were reduced with the UVES pipeline software (as described in more detail in Ritchey et al.~2011). The spectra have very high signal-to-noise ratios (S/N~$\sim$~1200--2000 per resolution element near 3875 \AA) and a resolution of approximately 3.5~km~s$^{-1}$ ($R=85,000$).

Figure~1 presents the VLT/UVES spectra for the four directions in the vicinity of the CN $B$$-$$X$ (0,~0) $R$(0) line. The grey line in each panel indicates the profile synthesis fit originally obtained by Ritchey et al.~(2011). In each case, it is apparent that there is extra absorption between the main $^{12}$CN and $^{13}$CN components that is not adequately accounted for by the synthetic spectrum. The expected isotope shift for the C$^{15}$N $R$(0) line from laboratory spectroscopy is +0.121 \AA{} or +9.4 km~s$^{-1}$ (Colin \& Bernath 2012; Sneden et al. 2014), which is consistent with the position of this weak additional feature (except in the case of HD~210121; see below). However, before we can claim a detection of the C$^{15}$N isotopologue, we must first consider the possibility that the extra absorption arises from an additional cloud contributing a weak $^{12}$CN $R$(0) component. To this end, we examined the $A$$-$$X$ (0,~0) lines of CH and CH$^+$, which were also covered by the UVES spectra (see Figures~2 and 3).

The first step in the analysis was to synthesize the CH lines, including as many weak additional absorption components beyond the main component as were necessary to provide a good fit. As in Ritchey et al.~(2011), we used the profile synthesis routine ISMOD (e.g., Sheffer et al.~2008), which determines best-fitting values for the velocities, $b$-values, and column densities of the absorption components through an iterative rms-minimizing procedure. The goodness of fit was evaluated by comparing the residuals to the noise in the continuum regions. After fitting the CH lines, we reexamined the CH$^+$ fits from Ritchey et al.~(2011) to determine if any additional components needed to be included. (Only toward HD~154368 did we see the need to include an additional weak CH$^+$ component.) For these CH$^+$ syntheses, the $^{12}$CH$^+$/$^{13}$CH$^+$ ratio was held fixed at 74.4, i.e, the mean value obtained by Ritchey et al.~(2011) from McDonald Observatory data. These fits are not significantly better or worse than ones where the ratio was allowed to vary, illustrating the difficulties one encounters when attempting to derive $^{12}$CH$^+$/$^{13}$CH$^+$ ratios from these VLT/UVES spectra (see Ritchey et al.~2011). Nonetheless, absorption from $^{13}$CH$^+$ had to be included at some level because in certain cases (i.e., toward HD~154368 and HD~169454) this absorption overlaps with weak additional $^{12}$CH$^+$ components blueward of the main components.

The next step was to use the component structure derived for CH and CH$^+$ (see Table~1) to determine which of the weak additional components might need to be included in the CN syntheses. In general, we retained only the strongest of the additional CH components if there was evidence of absorption at those velocities in the CN profile. In many cases, the components not included have $N$(CH$^+$)~$>$~$N$(CH), suggesting that the CH in those clouds is associated with CH$^+$ rather than with CN (e.g., Zsarg{\'o} \& Federman 2003). Toward HD~169454, there is a weak CH component at $v_{\mathrm{LSR}}$~=~+16.7~km~s$^{-1}$ that is shifted by +10.9~km~s$^{-1}$ with respect to the main component at $v_{\mathrm{LSR}}$~=~+5.8~km~s$^{-1}$. This shift is close to the expected isotope shift of +9.4 km~s$^{-1}$ for C$^{15}$N. However, we do not believe that the absorption seen in the CN profile at this velocity is due to an additional component of $^{12}$CN. The cloud in question has $N$(CH$^+$)/$N$(CH)~$>$~1.0, and $b$(CH)~$>$~$b$(CH$^+$), indicating that the CH in this case is CH$^+$-like rather than CN-like (Lambert et al.~1990; Pan et al.~2005). Such gas should have very little (if any) associated CN. None of the other sight lines have a CH component that in the CN profile could be mistaken for C$^{15}$N. Toward HD~210121, there is a weak CH component shifted by +8.5~km~s$^{-1}$ relative to the main component. However, there is no evidence for this component in the CN profile, and, moreover, C$^{15}$N is not significantly detected in this direction (see Section~3).

We have made several improvements to the analysis of the CN features in the present work compared to our previous analysis of these spectra (Ritchey et al.~2011). Most importantly, we have adopted an updated set of $f$-values and wavelengths for all of the relevant CN transitions, using data from the recently-published CN line lists (Brooke et al.~2014; Sneden et al.~2014). In addition, we now explicitly account for the splitting in the CN $R$(0) feature, which is a blend of the $R_1$(0) and $^{R}Q_{21}$(0) lines. We adopt a splitting of 0.0039~\AA{} or 0.30~km s$^{-1}$, with the former line having an $f$-value twice that of the latter (Federman et al.~1984; Brooke et al.~2014). The isotope shifts for the $^{13}$CN and C$^{15}$N $R$(0) features were held fixed at the values derived from laboratory spectroscopy. For $^{13}$CN, the expected isotope shift is +0.168 \AA{} or +13.0~km~s$^{-1}$ (Ram \& Bernath 2011; Sneden et al. 2014). The column densities and $b$-values of the main $^{12}$CN absorption components were determined through simultaneous fits to the $R$(0) lines of the (0,~0) and (1,~0) bands of the CN $B$$-$$X$ system. Spectra covering the CN (1,~0) band near 3580~\AA{} were obtained from archival VLT/UVES observations, which generally employed a different instrumental setup and have somewhat lower resolution than the spectra covering the (0,~0) band near 3875~\AA{} (see Ritchey et al.~2011).\footnote{For HD~73882, we use new observations of the CN (1,~0) band from VLT/UVES spectra acquired under program 092.C-0019 (PI: H.~Linnartz).} The $b$-values derived for the main $^{12}$CN absorption components were used to fit the corresponding $^{13}$CN and C$^{15}$N absorption features.

The final component structures obtained for CH$^+$, CH, and CN are given in Table~1, where the CH$^+$ and CN column densities refer only to the dominant isotopologue. Table~2 provides the column densities and resulting ratios of the CN isotopologues for the main absorption components. Uncertainties in the derived column densities account for both photon noise and errors in continuum placement. For the main absorption components, the column density uncertainties also include contributions related to the errors in the derived $b$-values. Corrections for optical depth are the dominant source of error for the very strong $^{12}$CN absorption lines seen here. Ritchey et al.~(2011) discussed the sensitivity of the $^{12}$CN column densities to very small changes in $b$. However, the $b$-value related uncertainties were not explicitly included in the errors reported in that paper. Here, we formally include column density uncertainties based on the errors in the $b$-values, which are very well constrained as a result of our simultaneous fits to the strong (0,~0) and weak (1,~0) $R$(0) lines (see Table~1).

\section{RESULTS AND DISCUSSION}
From a careful analysis of the component structure along each line of sight, we find that absorption from the C$^{15}$N isotopologue is confidently detected in three out of the four directions studied. For HD~210121, the extra absorption between the main $^{12}$CN and $^{13}$CN absorption features is mostly due to an additional component of $^{12}$CN, which is apparent in the CH profile as well. The $3\sigma$ upper limit on the equivalent width of the C$^{15}$N line in this direction is $W_{\lambda}$(C$^{15}$N)~$\lesssim0.20$~m\AA. For the other directions, we find C$^{15}$N equivalent widths of $0.399\pm0.057$~m\AA{} (HD~73882), $0.178\pm0.042$~m\AA{} (HD~154368), and $0.487\pm0.035$~m\AA{} (HD~169454), corresponding to $7.0\sigma$, $4.2\sigma$, and $13.9\sigma$ detections, respectively. The weighted mean C$^{14}$N/C$^{15}$N ratio for the three sight lines where C$^{15}$N is detected is $274\pm18$. Since the diffuse molecular clouds toward our target stars have relatively high kinetic temperatures ($T_{\mathrm{kin}}\gtrsim20$~K) and the visual extinctions are relatively low ($A_V\lesssim4$~mag), the C$^{14}$N/C$^{15}$N ratios we derive should not be affected by chemical or photochemical fractionation (see below). Our mean value should therefore be representative of the ambient $^{14}$N/$^{15}$N ratio in the local ISM.

In principle, two processes could drive the C$^{14}$N/C$^{15}$N ratio away from the ambient $^{14}$N/$^{15}$N ratio in interstellar clouds. Isotope selective photodissociation of N$_2$ could alter the atomic $^{14}$N/$^{15}$N ratio from which all other nitrogen-bearing molecules are produced.  However, the detailed analysis by Heays et al.~(2014) indicates that significant reductions in $^{14}$N/$^{15}$N, resulting from enhancements in $^{14}$N$_2$/$^{14}$N$^{15}$N, occur only at relatively large extinctions (i.e., $A_V\gtrsim4$~mag).\footnote{Since the interstellar cloud model of Heays et al.~(2014) employs a one-sided irradiation geometry, the extinctions in that paper need to be multiplied by two for comparison with the total line-of-sight extinctions toward our targets.} Our targets have line-of-sight extinctions in the range $A_V$~$\approx$~0.8--3.6~mag (e.g., Valencic et al.~2004; Rachford et al.~2009; see Table~2). Chemical fractionation (e.g., Hily-Blant et al.~2013b) involving ion-molecule reactions leading to nitrogen hydride ions, which are precursors in the formation of CN (e.g., Federman et al.~1994), requires gas temperatures of $\sim$10~K. However, the clouds probed by our targets have temperatures (from analyses of C$_2$ rotational excitation; e.g., Sonnentrucker et al.~2007) of $T_{\mathrm{kin}}\approx20\pm5$~K (HD~73882, HD~154368, and HD~169454) and $45\pm5$~K (HD~210121).

The $^{12}$CN/$^{13}$CN ratios for the sight lines studied here have been revised slightly compared to the ratios given in Ritchey et al.~(2011), as a result of the updated $f$-values that we adopt and the inclusion of absorption from the C$^{15}$N isotopologue, among other changes (see Section 2). However, the new results do not significantly alter the conclusions of Ritchey et al.~(2011). The weighted mean value of $^{12}$CN/$^{13}$CN from our new determinations for HD~73882, HD~154368, HD~169454, and HD~210121 is $69.6\pm1.6$. If we include results for the other sight lines analyzed by Ritchey et al.~(2011), based on the same set of $f$-values as used here, then the weighted mean $^{12}$CN/$^{13}$CN ratio we obtain is $64.3\pm1.5$. (The lower mean value for this larger sample results from the inclusion of several sight lines with markedly low $^{12}$CN/$^{13}$CN ratios; such ratios are presumably a consequence of selective photodissociation of CO, which can be important at relatively low values of $A_V$; e.g., Visser et al.~2009). In either case, we find that the mean $^{12}$CN/$^{13}$CN ratio for local diffuse molecular clouds is in good agreement with other estimates for the $^{12}$C/$^{13}$C ratio in the local ISM (e.g., Wilson 1999; Milam et al.~2005). Milam et al.~(2005), examining millimeter-wave observations of CN, CO, and H$_2$CO in dense molecular clouds, inferred a carbon isotopic ratio for the solar neighborhood of $68\pm15$ from a fit to the $^{12}$C/$^{13}$C gradient with Galactocentric distance. A similar correspondence is seen for the $^{14}$N/$^{15}$N ratio. Our mean value of C$^{14}$N/C$^{15}$N for local diffuse clouds ($274\pm18$) is in good agreement with the nitrogen isotopic ratio derived for the solar neighborhood by Adande \& Ziurys (2012) from their fit to the Galactic gradient in $^{14}$N/$^{15}$N ($290\pm40$).

The close correspondence between the $^{12}$C/$^{13}$C and $^{14}$N/$^{15}$N ratios that we derive for local diffuse molecular clouds (from the analysis of optical interstellar absorption lines) and those measured for dense molecular clouds in the solar neighborhood (from millimeter observations) indicates that these values accurately reflect the ambient ratios in the present-day local ISM. In contrast, the $^{12}$C/$^{13}$C and $^{14}$N/$^{15}$N ratios attributed to the Sun and/or the protosolar nebula are both higher than the present-day interstellar ratios, presumably because the ISM has been enriched with the products of stellar nucleosynthesis since the formation of the solar system some 4.6 Gyr ago. Marty et al.~(2011), reporting on solar wind data acquired by the \emph{Genesis} mission, measured a $^{14}$N/$^{15}$N ratio of $459\pm4$, from which they derived a value of $441\pm6$ for the protosolar nebula. Our mean value of C$^{14}$N/C$^{15}$N for the local ISM ($274\pm18$), when compared with the protosolar nebula value of $^{14}$N/$^{15}$N, implies that a $^{15}$N enrichment of $\sim$60\% has occurred over the last 4.6 Gyr. A comparable study of the $^{12}$C/$^{13}$C ratio from the \emph{Genesis} mission has yet to be reported (D.~S.~Burnett 2014, private communication). However, this ratio is obtainable for the solar photosphere from infrared CO lines: Scott et al.~(2006) reported $^{12}$C/$^{13}$C~=~$86.8^{+3.9}_{-3.7}$ and Ayres et al.~(2013) determined a ratio of $^{12}$C/$^{13}$C~=~$91.4\pm1.3$ using different simulations of the solar granulation. Since the local interstellar value of $^{12}$C/$^{13}$C is $\sim$70, this implies that a $^{13}$C enrichment of $\sim$30\% has occurred since the formation of the Sun.

Interpretation of the evolution of the $^{12}$C/$^{13}$C and $^{14}$N/$^{15}$N ratios requires identification of the principal sources responsible for the synthesis of these isotopes. A recent comprehensive model of Galactic chemical evolution, which sought to reproduce the isotopic ratios seen in the Sun and in other stars throughout the Galaxy, was presented by Kobayashi et al.~(2011). In constructing their model, Kobayashi et al.~consider all of the major sources of stellar nucleosynthesis, including stellar winds, AGB and super-AGB stars, Type Ia supernovae, core-collapse supernovae (including hypernovae), and pair-instability supernovae, and describe recipes for the initial mass function and the star formation histories. The $^{12}$C/$^{13}$C ratio predicted by the Kobayashi et al.~model at [Fe/H]~=~0 is $\sim$89, in very good agreement with the solar ratio. However, the predicted $^{14}$N/$^{15}$N ratio at [Fe/H]~=~0 is $\sim$2500, which is significantly larger than the ratio inferred for the protosolar nebula. Kobayashi et al.~remark that this discrepancy for $^{14}$N/$^{15}$N is most likely due to the fact that nova nucleosynthesis is not included in their model. The GCE models of Romano \& Matteucci (2003, 2005), on the other hand, do include contributions from novae, along with the standard sources of stellar nucleosynthesis. These authors adjusted the contributions from novae so that the solar isotopic ratios (i.e., $^{12}$C/$^{13}$C, $^{14}$N/$^{15}$N, and $^{16}$O/$^{17}$O) were reproduced at the appropriate time. With this adjustment, their GCE models account well for the observed decreases in the $^{12}$C/$^{13}$C and $^{14}$N/$^{15}$N ratios over the 4.6 Gyr between the formation of the Sun and the present. The yields from novae for $^{13}$C, $^{15}$N, and $^{17}$O are predicted to be dependent on the mass of the C-O or O-Ne white dwarf and the degree of mixing between the accreted envelope and the surface of the white dwarf on which the explosion occurs (Jos\'{e} \& Hernanz 1998). The key nuclear process is the hot CNO cycle. If, as Romano \& Matteucci (2005) assert, ``novae are at present the only candidates for producing $^{15}$N in a galaxy,'' future studies of $^{15}$N in gas and stars in the Milky Way and other galaxies are to be encouraged.

\acknowledgments

We thank Peter Bernath and Don Burnett for helpful comments related to this work. The research presented here was supported by NASA grant NNX10AD80G (S.R.F.), and by the Kenilworth Fund of the New York Community Trust (A.M.R.). D.L.L. thanks the Robert A.~Welch Foundation of Houston, TX for support through grant F-634.

\clearpage

\begin{figure}
\centering
\includegraphics[width=1.0\textwidth]{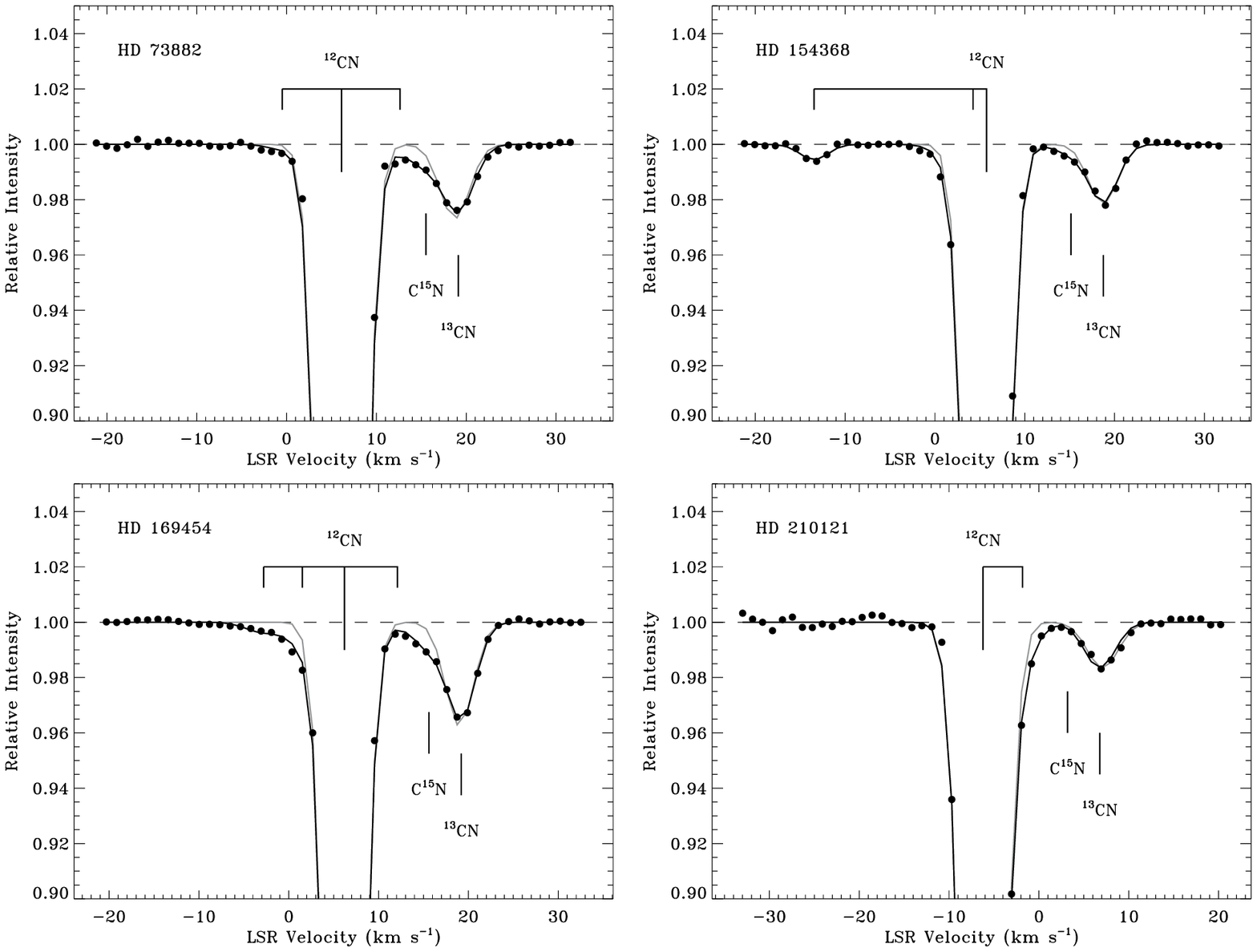}
\caption[]{Profile synthesis fits to the CN $B$$-$$X$ (0,~0) $R$(0) lines toward HD~73882, HD~154368, HD~169454, and HD~210121. The reference wavelength for the velocity scale is $\lambda_0=3874.602$~\AA{}. Synthetic profiles are shown as solid lines passing through data points that represent the observed spectra. The grey line represents the fit from Ritchey et al.~(2011), while the black line shows the fit obtained in this work. The position of the main absorption component in $^{12}$CN is indicated by a longer tick mark, while shorter tick marks give the positions of weak additional $^{12}$CN components included in the fit. The positions of the main $^{13}$CN and C$^{15}$N components are also indicated.}
\end{figure}

\clearpage

\begin{figure}
\centering
\includegraphics[width=1.0\textwidth]{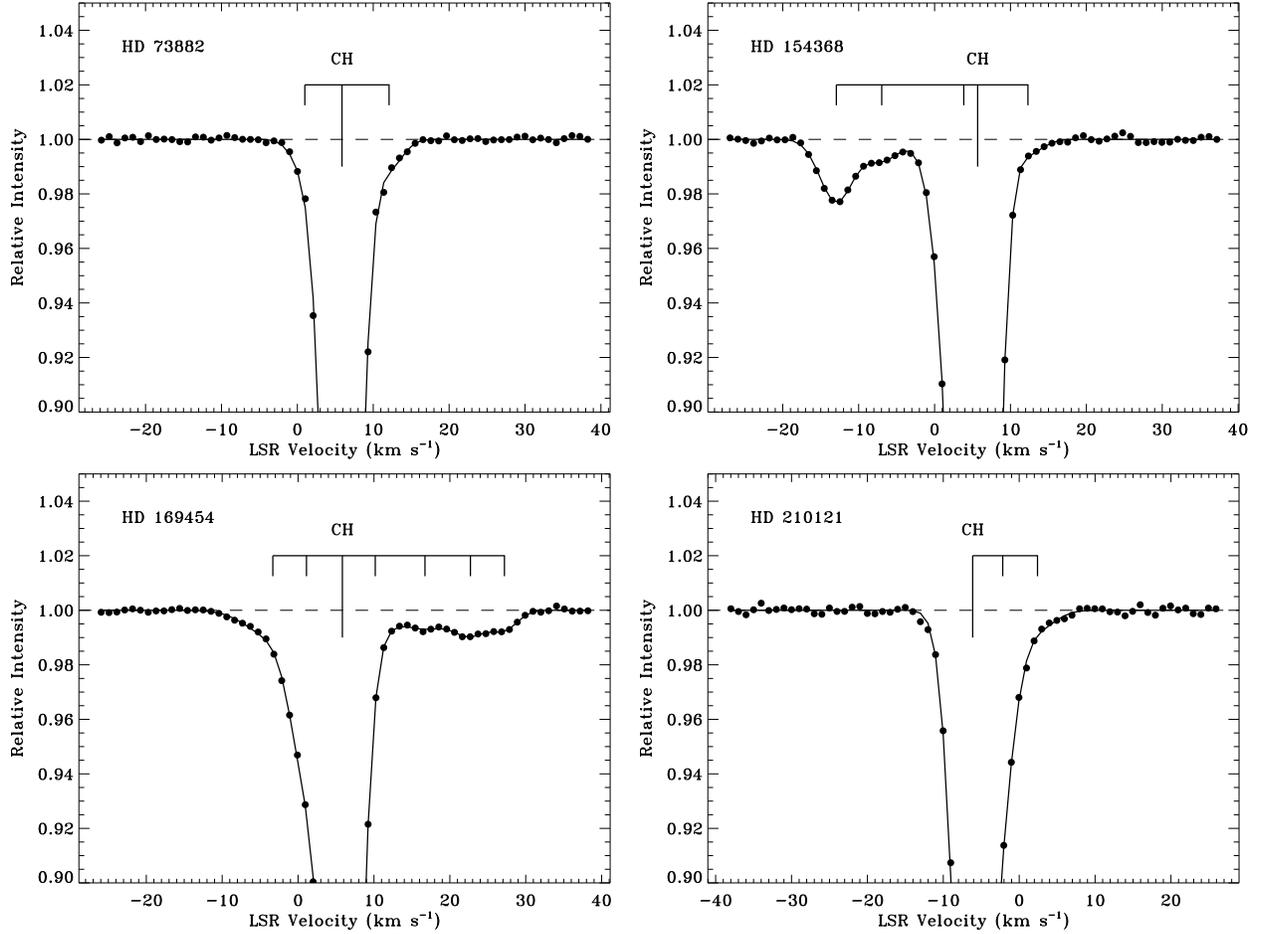}
\caption[]{Profile synthesis fits to the CH $A$$-$$X$ (0,~0) lines toward HD~73882, HD~154368, HD~169454, and HD~210121 ($\lambda_0=4300.313$~\AA{}). Plotting symbols are the same as in Figure~1. The position of the main CH absorption component is indicated by a longer tick mark, while shorter tick marks give the positions of weak additional components included in the fit.}
\end{figure}

\clearpage

\begin{figure}
\centering
\includegraphics[width=1.0\textwidth]{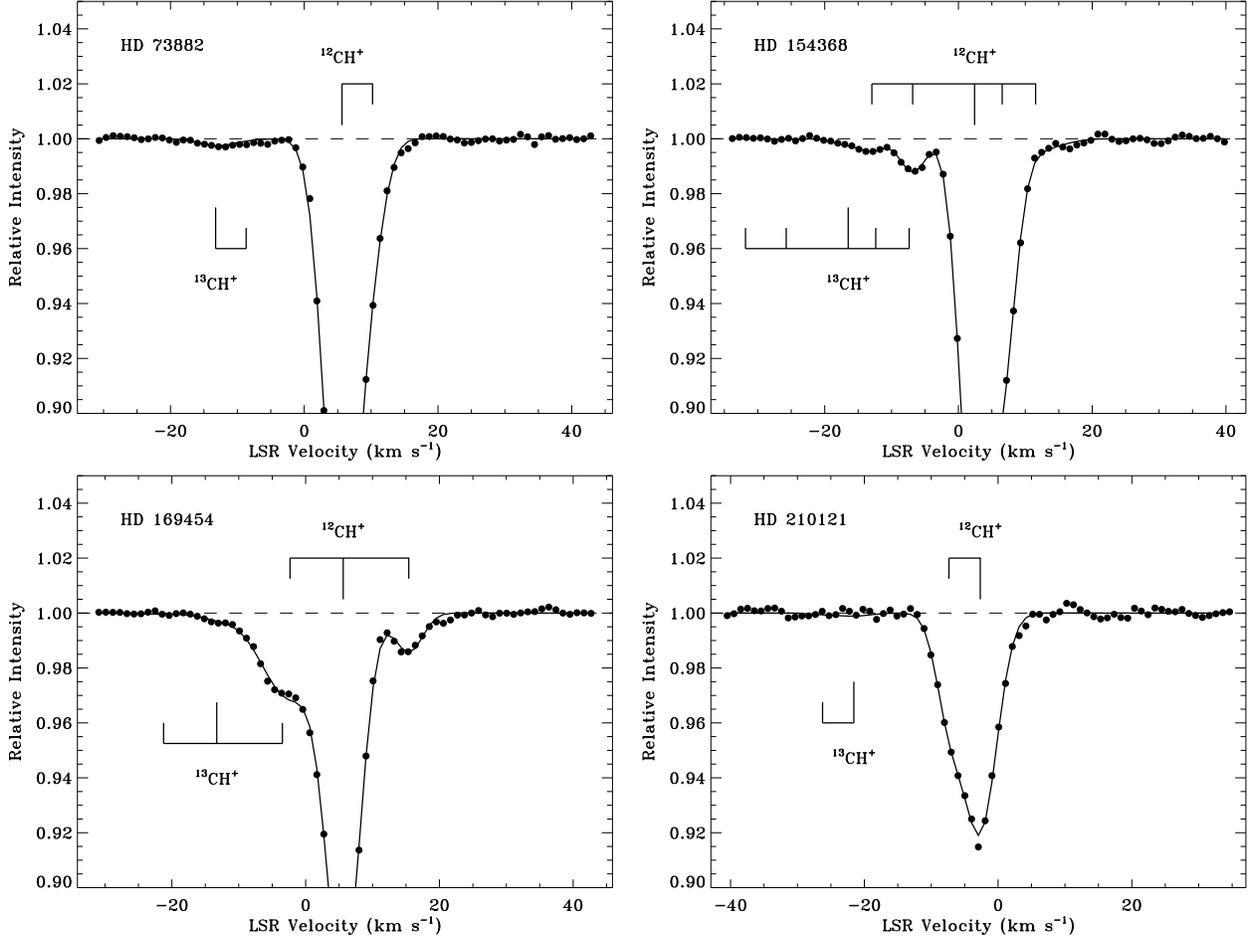}
\caption[]{Profile synthesis fits to the CH$^+$ $A$$-$$X$ (0,~0) lines toward HD~73882, HD~154368, HD~169454, and HD~210121 ($\lambda_0=4232.548$~\AA{}). Plotting symbols are the same as in Figure~1. The positions of the main absorption components in $^{12}$CH$^+$ and $^{13}$CH$^+$ are indicated by longer tick marks, while shorter tick marks give the positions of weak additional components included in the fit. The two CH$^+$ isotopologues are assumed to have identical component structure and to exhibit a $^{12}$CH$^+$/$^{13}$CH$^+$ ratio of 74.4 (see the text).}
\end{figure}

\clearpage

\begin{deluxetable}{lccccccccccc}
\rotate
\tablecolumns{12}
\tablewidth{0pt}
\tabletypesize{\footnotesize}
\tablecaption{CH$^+$, CH, and CN Component Structure}
\tablehead{\colhead{Star} & \multicolumn{3}{c}{CH$^+$} & \colhead{} & \multicolumn{3}{c}{CH} & \colhead{} & \multicolumn{3}{c}{CN} \\
\cline {2-4} \cline{6-8} \cline{10-12} \\
\colhead{} & \colhead{$v_{\mathrm{LSR}}$} & \colhead{$N$} & \colhead{$b$} & \colhead{} & \colhead{$v_{\mathrm{LSR}}$} & \colhead{$N$} & \colhead{$b$} & \colhead{} & \colhead{$v_{\mathrm{LSR}}$} & \colhead{$N$} & \colhead{$b$\tablenotemark{a}} \\
\colhead{} & \colhead{(km s$^{-1}$)} & \colhead{($10^{12}$ cm$^{-2}$)} & \colhead{(km s$^{-1}$)} & \colhead{} & \colhead{(km s$^{-1}$)} & \colhead{($10^{12}$ cm$^{-2}$)} & \colhead{(km s$^{-1}$)} & \colhead{} & \colhead{(km s$^{-1}$)} & \colhead{($10^{12}$ cm$^{-2}$)} & \colhead{(km s$^{-1}$)} }
\startdata
HD~73882 & \ldots & \ldots & \ldots &&       \phn+1.0 &  $\phn0.71\pm0.04$ & 0.2 && \phn$-$0.5 &  $\phn0.02\pm0.01$ & 0.2 \\
         &   \phn+5.6 & $19.58\pm0.09$ & 2.8 &&  \phn+5.9 & $33.44\pm0.44$ & 1.4 &&   \phn+6.1 & $20.70\pm0.83$ & $0.979\pm0.022$ \\
         &  +10.2 &  $\phn2.39\pm0.07$ & 1.9 && +12.1 &  $\phn0.79\pm0.04$ & 0.2 &&  +12.6 &  $\phn0.02\pm0.01$ & 0.2 \\
HD~154368 & $-$12.9 &  $\phn0.20\pm0.04$ & 0.5 && $-$12.9 &  $\phn2.15\pm0.06$ & 1.8 && $-$13.4 &  $\phn0.06\pm0.01$ & 0.2 \\
          &  \phn$-$6.8 &  $\phn0.93\pm0.05$ & 1.4 &&  \phn$-$7.0 &  $\phn0.75\pm0.06$ & 1.9 && \ldots & \ldots & \ldots \\
          &    \phn+2.4 & $14.40\pm0.08$ & 2.0 &&    \phn+3.8 & $15.89\pm0.08$ & 2.6 &&    \phn+4.2 &  $\phn0.26\pm0.01$ & 2.6 \\
          &    \phn+6.5 &  $\phn6.98\pm0.06$ & 1.9 &&    \phn+5.7 & $45.75\pm1.19$ & 1.1 &&    \phn+5.7 & $17.78\pm0.47$ & $0.641\pm0.008$ \\
          &   +11.5 &  $\phn0.62\pm0.10$ & 4.5 &&   +12.3 &  $\phn0.43\pm0.05$ & 1.1 && \ldots & \ldots & \ldots \\
HD~169454 & \phn$-$2.3 &  $\phn5.36\pm0.09$ & 5.5 && \phn$-$3.3 &  $\phn1.40\pm0.07$ & 4.0 && \phn$-$2.8 &  $\phn0.05\pm0.01$ & 1.6 \\
          & \ldots & \ldots & \ldots &&        \phn+1.1 &  $\phn4.95\pm0.04$ & 1.6 &&   \phn+1.5 &  $\phn0.09\pm0.01$ & 0.6 \\
          &   \phn+5.6 & $14.17\pm0.06$ & 2.7 &&   \phn+5.8 & $33.67\pm0.50$ & 1.3 &&   \phn+6.2 & $29.47\pm0.80$ & $0.464\pm0.004$ \\
          & \ldots & \ldots & \ldots &&       +10.2 &  $\phn0.67\pm0.04$ & 2.0 &&  +12.1 &  $\phn0.02\pm0.01$ & 0.8 \\
          &  +15.4 &  $\phn1.20\pm0.04$ & 1.9 &&  +16.7 &  $\phn0.85\pm0.06$ & 3.3 && \ldots & \ldots & \ldots \\
          & \ldots & \ldots & \ldots &&       +22.7 &  $\phn0.80\pm0.04$ & 1.6 && \ldots & \ldots & \ldots \\
          & \ldots & \ldots & \ldots &&       +27.2 &  $\phn0.52\pm0.03$ & 0.2 && \ldots & \ldots & \ldots \\
HD~210121 & \phn$-$7.3 & $\phn2.84\pm0.10$ & 1.6 && \phn$-$6.1 & $26.14\pm0.22$ & 1.5 && \phn$-$6.2 & $14.03\pm0.32$ & $0.785\pm0.014$ \\
          & \phn$-$2.7 & $\phn8.30\pm0.12$ & 2.6 && \phn$-$2.2 &  $\phn4.13\pm0.08$ & 0.8 && \phn$-$1.8 &  $\phn0.14\pm0.01$ & 0.4 \\
          & \ldots & \ldots & \ldots &&       \phn+2.4 &  $\phn0.67\pm0.10$ & 2.0 && \ldots & \ldots & \ldots \\
\enddata
\tablenotetext{a}{The $b$-values of the main CN absorption components (those with error bars) were determined through simultaneous fits to the $R$(0) lines of the (0,~0) and (1,~0) bands.}
\end{deluxetable}

\begin{deluxetable}{lccccccc}
\tablecolumns{8}
\tablewidth{0pt}
\tabletypesize{\footnotesize}
\tablecaption{Column Densities and Ratios of CN Isotopologues}
\tablehead{\colhead{Star} & \colhead{$A_V$\tablenotemark{a}} & \colhead{$N$($^{12}$C$^{14}$N)} & \colhead{$N$($^{13}$C$^{14}$N)} & \colhead{$N$($^{12}$C$^{15}$N)} & \colhead{$^{12}$CN/$^{13}$CN} & \colhead{C$^{14}$N/C$^{15}$N} \\
\colhead{} & \colhead{(mag)} & \colhead{($10^{12}$ cm$^{-2}$)} & \colhead{($10^{11}$ cm$^{-2}$)} & \colhead{($10^{11}$ cm$^{-2}$)} & \colhead{} & \colhead{} }
\startdata
HD~73882  & $2.42\pm0.14$ & $20.70\pm0.83$ & $3.02\pm0.13$ & $0.88\pm0.13$ & $68.6\pm4.0$ & $234\pm35$\phn \\
HD~154368 & $2.46\pm0.16$ & $17.78\pm0.47$ & $2.49\pm0.09$ & $0.39\pm0.09$ & $71.4\pm3.3$ & $452\pm107$ \\
HD~169454 & $3.64\pm0.17$ & $29.47\pm0.80$ & $4.23\pm0.08$ & $1.04\pm0.08$ & $69.7\pm2.3$ & $283\pm22$\phn \\
HD~210121 & $0.79\pm0.10$ & $14.03\pm0.32$ & $2.11\pm0.15$ & $<0.45$       & $66.5\pm4.9$ & $>312$ \\
\enddata
\tablenotetext{a}{Weighted mean of the values listed in Valencic et al.~(2004) and Rachford et al.~(2009), except in the case of HD~169454, for which $A_V$ is available only from Valencic et al.~(2004).}
\end{deluxetable}

\end{document}